\documentclass[a4paper,12pt]{article}
\usepackage{graphicx}
\usepackage{graphics}
\usepackage{epsf}
\begin{document}
\newcommand{\be}{\begin{equation}}
\newcommand{\beq}{\begin{equation}}
\newcommand{\eeq}{\end{equation}}
\newcommand{\ee}{\end{equation}}

\newcommand{\beqn}{\begin{eqnarray}}
\newcommand{\eeqn}{\end{eqnarray}}
\newcommand{\bea}{\begin{eqnarray}}
\newcommand{\ena}{\end{eqnarray}}
\newcommand{\ra}{\rightarrow}
\newcommand{\susy}{{{\cal SUSY}$\;$}}
\newcommand{\su}{$ SU(2) \times U(1)\,$}

\newcommand{\gag}{$\gamma \gamma$ }
\newcommand{\gagt}{\gamma \gamma }
\newcommand{\gam}{\gamma \gamma }
\def\W{{\mbox{\boldmath $W$}}}
\def\B{{\mbox{\boldmath $B$}}}
\def\V{{\mbox{\boldmath $V$}}}
\newcommand{\np}{Nucl.\,Phys.\,}
\newcommand{\pl}{Phys.\,Lett.\,}
\newcommand{\pr}{Phys.\,Rev.\,}
\newcommand{\prl}{Phys.\,Rev.\,Lett.\,}
\newcommand{\prep}{Phys.\,Rep.\,}
\newcommand{\zp}{Z.\,Phys.\,}
\newcommand{\sovjnp}{{\em Sov.\ J.\ Nucl.\ Phys.\ }}
\newcommand{\nuclinst}{{\em Nucl.\ Instrum.\ Meth.\ }}
\newcommand{\annp}{{\em Ann.\ Phys.\ }}
\newcommand{\intjmp}{{\em Int.\ J.\ of Mod.\  Phys.\ }}

\newcommand{\eps}{\epsilon}
\newcommand{\mw}{M_{W}}
\newcommand{\mww}{M_{W}^{2}}
\newcommand{\mwmw}{M_{W}^{2}}
\newcommand{\mhmh}{M_{H}^2}
\newcommand{\mz}{M_{Z}}
\newcommand{\mzz}{M_{Z}^{2}}

\newcommand{\cw}{\cos\theta_W}
\newcommand{\sw}{\sin\theta_W}
\newcommand{\tw}{\tan\theta_W}
\def\cww{\cos^2\theta_W}
\def\sww{s^2_W}
\def\tww{\tan^2\theta_W}

\newcommand{\epm}{$e^{+} e^{-}\;$}
\newcommand{\epemt}{$e^{+} e^{-}\;$}
\newcommand{\epem}{e^{+} e^{-}\;}
\newcommand{\ememt}{$e^{-} e^{-}\;$}
\newcommand{\emem}{e^{-} e^{-}\;}

\newcommand{\lra}{\leftrightarrow}
\newcommand{\tr}{{\rm Tr}}
\def\ls1{{\not l}_1}
\newcommand{\cms}{centre-of-mass\hspace*{.1cm}}

\newcommand{\ie}{{\em i.e.}}
\newcommand{\cm}{{{\cal M}}}
\newcommand{\cl}{{{\cal L}}}
\newcommand{\cd}{{{\cal D}}}
\newcommand{\cv}{{{\cal V}}}
\def\slashc{c\kern -.400em {/}}
\def\slashp{p\kern -.400em {/}}
\def\slashL{L\kern -.450em {/}}
\def\slashcl{\cl\kern -.600em {/}}
\def\Ww{{\mbox{\boldmath $W$}}}
\def\B{{\mbox{\boldmath $B$}}}
\def\noi{\noindent}
\def\nn{\noindent}
\def\sm{${\cal{S}} {\cal{M}}\;$}
\def\smn{${\cal{S}} {\cal{M}}$}
\def\nph{${\cal{N}} {\cal{P}}\;$}
\def\sb{$ {\cal{S}}  {\cal{B}}\;$}
\def\ssb{${\cal{S}} {\cal{S}}  {\cal{B}}\;$}
\def\ssbe{{\cal{S}} {\cal{S}}  {\cal{B}}}
\def\cviol{${\cal{C}}\;$}
\def\pviol{${\cal{P}}\;$}
\def\cpviol{${\cal{C}} {\cal{P}}\;$}

\newcommand{\sinsq}{\sin^2\theta}
\newcommand{\cossq}{\cos^2\theta}
\newcommand{\yt}{y_\theta}

\def\sinb{\sin\beta}
\def\cosb{\cos\beta}
\def\sinbb{\sin (2\beta)}
\def\cosbb{\cos (2 \beta)}
\def\tgb{\tan \beta}
\def\tgbt{$\tan \beta\;\;$}
\def\tgbsq{\tan^2 \beta}
\def\sinal{\sin\alpha}
\def\cosal{\cos\alpha}
\def\stop{\tilde{t}}
\def\sto{\tilde{t}_1}
\def\stt{\tilde{t}_2}
\def\stl{\tilde{t}_L}
\def\str{\tilde{t}_R}
\def\msto{m_{\sto}}
\def\mstosq{m_{\sto}^2}
\def\mstt{m_{\stt}}
\def\msttsq{m_{\stt}^2}
\def\mt{m_t}
\def\mtsq{m_t^2}
\def\sint{\sin\theta_{\stop}}
\def\sintt{\sin 2\theta_{\stop}}
\def\cost{\cos\theta_{\stop}}
\def\sintsq{\sin^2\theta_{\stop}}
\def\costsq{\cos^2\theta_{\stop}}
\def\mqtt{\M_{\tilde{Q}_3}^2}
\def\mutt{\M_{\tilde{U}_{3R}}^2}
\def\sbottom{\tilde{b}}
\def\sbo{\tilde{b}_1}
\def\sbt{\tilde{b}_2}
\def\sbl{\tilde{b}_L}
\def\sbr{\tilde{b}_R}
\def\msbo{m_{\sbo}}
\def\msbosq{m_{\sbo}^2}
\def\msbt{m_{\sbt}}
\def\msbtsq{m_{\sbt}^2}
\def\mt{m_t}
\def\mtsq{m_t^2}
\def\selectron{\tilde{e}}
\def\seo{\tilde{e}_1}
\def\set{\tilde{e}_2}
\def\sel{\tilde{e}_L}
\def\ser{\tilde{e}_R}
\def\mseo{m_{\seo}}
\def\mseosq{m_{\seo}^2}
\def\mset{m_{\set}}
\def\msetsq{m_{\set}^2}
\def\msel{m_{\sel}}
\def\mser{m_{\ser}}
\def\me{m_e}
\def\mesq{m_e^2}
\def\snu{\tilde{\nu}}
\def\snue{\tilde{\nu_e}}
\def\set{\tilde{e}_2}
\def\snul{\tilde{\nu}_L}
\def\msnue{m_{\snue}}
\def\msnuesq{m_{\snue}^2}
\def\smuon{\tilde{\mu}}
\def\smul{\tilde{\mu}_L}
\def\smur{\tilde{\mu}_R}
\def\msmul{m_{\smul}}
\def\msmulsq{m_{\smul}^2}
\def\msmur{m_{\smur}}
\def\msmursq{m_{\smur}^2}
\def\stau{\tilde{\tau}}
\def\stauo{\tilde{\tau}_1}
\def\staut{\tilde{\tau}_2}
\def\staul{\tilde{\tau}_L}
\def\staur{\tilde{\tau}_R}
\def\mstauo{m_{\stauo}}
\def\mstauosq{m_{\stauo}^2}
\def\mstaut{m_{\staut}}
\def\mstautsq{m_{\staut}^2}
\def\mtau{m_\tau}
\def\mtausq{m_\tau^2}
\def\gluino{\tilde{g}}
\def\mgluino{m_{\tilde{g}}}
\def\mchi{m_\chi^+}
\def\neuto{\tilde{\chi}_1^0}
\def\mneuto{m_{\tilde{\chi}_1^0}}
\def\neutt{\tilde{\chi}_2^0}
\def\mneutt{m_{\tilde{\chi}_2^0}}
\def\neutth{\tilde{\chi}_3^0}
\def\mneutth{m_{\tilde{\chi}_3^0}}
\def\neutf{\tilde{\chi}_4^0}
\def\mneutf{m_{\tilde{\chi}_4^0}}
\def\chargop{\tilde{\chi}_1^+}
\def\mchargo{m_{\tilde{\chi}_1^+}}
\def\chargtp{\tilde{\chi}_2^+}
\def\mchargt{m_{\tilde{\chi}_2^+}}
\def\chargom{\tilde{\chi}_1^-}
\def\chargtm{\tilde{\chi}_2^-}
\def\bino{\tilde{b}}
\def\wino{\tilde{w}}
\def\photino{\tilde{\gamma}}
\def\zino{tilde{z}}
\def\sdowno{\tilde{d}_1}
\def\sdownt{\tilde{d}_2}
\def\sdownl{\tilde{d}_L}
\def\sdownr{\tilde{d}_R}
\def\supo{\tilde{u}_1}
\def\supt{\tilde{u}_2}
\def\supl{\tilde{u}_L}
\def\supr{\tilde{u}_R}
\def\mh{m_h}
\def\mht{m_h^2}
\def\MH{M_H}
\def\MHt{M_H^2}
\def\MA{M_A}
\def\MAt{M_A^2}
\def\MHp{M_H^+}
\def\MHm{M_H^-}
\def\gstar{g_*^{1/2}}
\def\bbar{b\overline{b}}
\def\ttbar{t\overline{t}}
\def\ccbar{c\overline{c}}
\def\micro{{\tt micrOMEGAs}}
\def\darksusy{{\tt DarkSusy}}
\def\comphep{{\tt CompHEP}}
\def\isasugra{{\tt ISASUGRA/Isajet}}
\def\isajet{{\tt Isajet}~}
\baselineskip=18pt

\bibliographystyle{unsrt}
\begin{titlepage}
\def\baselinestretch{1.2}
\topmargin     -0.25in

\vspace*{\fill}
\begin{center}
{\large {\bf micr\Large{OMEGA}s: A program for calculating the relic density
in the MSSM
 }} \vspace*{0.5cm}

\begin{tabular}[t]{c}

{\bf G.~B\'elanger$^{1}$, F.~Boudjema$^{1}$,  A. Pukhov$^{2}$,
A. Semenov$^{1}$}
 \\
\\
\\
{\it 1. Laboratoire de Physique Th\'eorique}
{\large LAPTH}
\footnote{URA 14-36 du CNRS, associ\'ee  \`a
l'Universit\'e de Savoie.}\\
 {\it Chemin de Bellevue, B.P. 110, F-74941 Annecy-le-Vieux,
Cedex, France.}\\

{\it 2. Skobeltsyn Institute of Nuclear Physics, 
Moscow State University} \\ {\it Moscow 119992,
Russia }\\

\end{tabular}
\end{center}

\centerline{ {\bf Abstract} }
\baselineskip=14pt
\noindent
 {\small We present a code that   calculates the relic density of the lightest supersymmetric
 particle (LSP) in the minimal
 supersymmetric standard model. All tree-level processes for the annihilation of the
 LSP are included as well as all possible 
 coannihilation processes with  neutralinos, charginos, sleptons, 
 squarks and gluinos. In all we have included over 2800
 processes not counting charged conjugate states.
 The cross-sections extracted from {\tt CompHEP} are calculated exactly. Relativistic
 formulae for the thermal average are used and care is taken to handle poles and
 thresholds by adopting specific integration routines. 
 The input parameters can be either the soft SUSY parameters in a general MSSM
 or the five parameters of a SUGRA model. A link with {\tt ISASUGRA/ISAJET} allows to
 calculate the parameters of the MSSM at the weak scale for an input at the GUT
 scale. The Higgs masses are
 calculated with {\tt FeynHiggsFast}. Routines calculating $(g-2)_\mu$ and 
 $b\to s\gamma$ are also included.}
\vspace*{\fill}


\vspace*{0.1cm}
\rightline{LAPTH-881/01}
\rightline{{\large  Dec. 2001}}
\end{titlepage}
\baselineskip=18pt

\setcounter{section}{0}
\setcounter{subsection}{0}
\setcounter{equation}{0}

\def\thesubsection {\thesection.\arabic{subsection}}
\def\theequation{\thesection.\arabic{equation}}

\section{Introduction}

One of the very attractive arguments in  favour of supersymmetry(SUSY) is that it provides 
a natural solution to the dark matter
problem. In R-parity conserving supersymmetric models there exists a neutral stable particle,
 the lightest
supersymmetric particle(LSP), which could constitute the cold  dark matter in the universe. 
As there are strong constraints on stable charged particles, only if the LSP 
is a neutralino or a sneutrino could SUSY provide a suitable dark matter candidate.
Cosmological constraints however as well as  direct searches disfavour the sneutrino
\cite{sneutrino} leaving the neutralino as the preferred
candidate for dark matter.  

The contribution of neutralinos to the relic density is however very model dependent 
and varies
by several orders of magnitude over the whole allowed parameter space of the MSSM.
The relic density then
imposes stringent constraints on the
parameters of the MSSM often favouring solutions with light supersymmetric particles.
It
 has been known for some time that  coannihilation processes where the LSP interacts 
with only
slightly heavier sparticles can  significantly reduce  the estimate of the relic density,
leading to acceptable values even with a rather heavy sparticle spectrum.
In principle these coannihilations can occur with any 
supersymmetric particle\cite{Griest}.  
The importance of the coannihilation channels were emphasized before both for
gauginos \cite{Yamaguchi,EdsjoGondolo}, sleptons\cite{Ellis-coann,GLP} or stops
\cite{abdelstop,Ellisstop}. Here, {\em ALL} channels will be included.
In some regions of the parameter space the neutralinos annihilate so fast  that 
they cannot constitute the only source of dark matter.   This happens 
for example when the
masses are such that the neutralinos can  annihilate through a 
s-channel Higgs resonance 
\cite{Ellis-Higgs,DreesNojiri} or a s-channel Z resonance\cite{g-2nous}.
In this context the inclusion of the relic density constraints, by giving a handle on the
supersymmetric particle spectrum, has important consequences both for
studies of SUSY at colliders and in astroparticle experiments.

There exist many calculations of the relic density in supersymmetry, using various 
approximations both in the
evaluation of cross-sections and in solving the density
equation \cite{Ellis-Higgs},\cite{neutdriver},\cite{Darksusy},\cite{Leszek},\cite{Torino},
\cite{BaerBrhlik}. 
Among these, {\tt Neutdriver}\cite{neutdriver}and {\tt DarkSusy}\cite{Darksusy} are publicly available.
Our purpose is to provide a tool that evaluates with high accuracy 
the annihilation cross-sections even in regions near poles and thresholds,  
that is both  flexible and upgradable  and that goes beyond {\tt DarkSusy}
 as far as the
calculation of the relic density is concerned.
 The main characteristics of this program, called \micro, are
\begin {itemize}
\item{} Complete tree-level matrix elements for all subprocesses
\item{} Includes all coannihilation channels with neutralinos, charginos, 
sleptons, squarks and gluinos. 
\item{} Loop-corrected Higgs masses and widths
\item{} Speed of calculation
\end{itemize}
All calculations of cross-sections are based on \comphep\cite{comphep}, an automatic program 
for the evaluation of
tree-level Feynman diagrams. We follow the method proposed by \cite{GondoloGelmini} for the calculation of
the relic density with its generalization to the case of coannihilations
\cite{EdsjoGondolo}. 
 We still rely  on approximations for the solution of the
relic density equations and the determination of the freeze-out temperature,
 since this allows to significantly increase the speed of the program
and proves to be  very
useful when scanning over a large parameter space. This, together with the fact that we have included
sfermion coannihilation channels as well as one-loop corrections to the Higgs width
constitute the main differences with {\tt DarkSUSY}.
Although 
we will generally assume that the neutralino is the LSP,
\micro~  can be used to compute the relic density with any supersymmetric
particle as  the LSP, in particular the sneutrino.
This is because all (co-)annihilation of any pairs of supersymmetric particles
into any pairs of standard model or Higgs particles
are included.

The program for the relic density calculation described here
is contained in a package that lets the user
 choose between  weak scale parameters or parameters of SUGRA models as input parameters.
The latter is achieved  through a link with \isasugra\cite{ISASUGRA}. The 
calculation of the  Higgs masses are done with {\tt FeynHiggsFast}\cite{FeynHiggs}. 
Loop QCD corrections
to the Higgs partial widths into fermions are extracted
from  {\tt HDECAY}\cite{HDECAY}. In addition we provide subroutines that calculate
various constraints on the MSSM parameters: direct limits from
colliders, $\Delta\rho$,
$b\to  s\gamma$ and $(g-2)_\mu$. 
All these constraints can be updated or replaced easily.

 
The total number  of processes  which can 
contribute to the relic density exceeds 2800.
However, due to a strong Boltzmann suppression factor, only  processes with  SUSY
particles close in mass to the LSP are relevant for the calculation.
Therefore in most cases 
 only a small fraction of the available processes are needed.
 In principle,  compilation of the full
set of subprocesses is  possible,
but such a program would be huge and could not be  distributed easily.
  To avoid this problem, we include in  our package the program
\comphep\cite{comphep}  which generates, while running, the  
subprocesses needed for a given set of  MSSM parameters.
The generated code  is linked during the run to the main
program and executed. The corresponding ``shared"  library  is
stored on the user disk space and is acessible for all subsequent
calls, thus
each process is generated and compiled only once.
Such approach can be realized only on 
Unix platforms which support dynamic linking.

The paper is organized as follows.
After we  summarize the important equations for the calculation of the relic density,
we give a short description of the parameters of the supersymmetric model.
A description of the package follows.
Finally we present  some results and comparisons with another program in the public domain, 
\darksusy.

\section{Calculation of the relic density}
The relic density at present in units of the critical density,
$\rho_{\rm crit}=3H^2/8\pi G$,  can be
expressed as
\beqn
\Omega_{\tilde{\chi}_1^0}=\frac{m_{\tilde{\chi}_1^0} n_{\tilde{\chi}_1^0}}{\rho_{{\rm
crit}}}=\frac{m_{\tilde{\chi}_1^0} s_0 Y_0}{\rho_{{\rm crit}}}
\eeqn

\noi
where $m_{\tilde{\chi}_1^0}$ is the mass of the lightest neutralino, $H=100 h ~ {\rm km} s^{-1}
Mpc^{-1}$ is the Hubble constant and $G$ Newton constant.
The entropy conservation assumption makes it easier to work
with the abundance $Y$ rather than the number density, $n_{\tilde{\chi}_1^0}$.
$s_0=s(T_0)$ defines  today's entropy to be evaluated at $T_0=2.726 K
$, the temperature of the microwave background, with
\beqn
s(T)=h_{\rm eff}(T)\frac{2 \pi^2}{45}T^3
\eeqn
where $h_{\rm eff}$ is a function that depends slowly on the temperature
$T$ \cite{Srednicki}. The code
in fact returns the relic density
\beqn
\label{omegah}
\Omega_{\tilde{\chi}_1^0} h^2=2.755 \times 10^8 \frac{m_{\tilde{\chi}_1^0}}{GeV} Y_0
\eeqn

One thus needs to find $Y_0=Y (T=T_0)$.  The most complete
formulae for the calculation of $Y(T)$ were presented in
\cite{GondoloGelmini,EdsjoGondolo} and we will follow their approach rather
closely.

The evolution equation of $Y$ is
\begin{equation}
   \frac{dY}{dT}= \sqrt{\frac{\pi  g_*(T) }{45G}} <\sigma v>(Y^2-Y_{eq}^2)
    \label{dydt}
\end{equation}
$g_*(T)$  is a degrees of
freedom parameter derived from the thermodynamics describing the state of the
universe \cite{Srednicki,OliveSteigman} and
$Y_{eq}=Y_{eq}(T)$ represents the thermal equilibrium abundance
\begin{equation}
Y_{eq}(T)=\frac{45}{4\pi^4h_{eff}(T)} \sum\limits_i
g_i\frac{m_i^2}{T^2} K_2(\frac{m_i}{T})
\end{equation}
where we sum over all supersymmetric particles $i$ with mass $m_i$
and $g_i$ degrees of freedom. $K_n$ is the modified Bessel
function of the second kind of order $n$ \cite{K}. Note that $Y_{eq}$ falls rather
rapidly as the temperature decreases. 
  $<\sigma v>$ is the relativistic thermally averaged
annihilation cross-section
\begin{equation}
       <\sigma v>=  \frac{ \sum\limits_{i,j}g_i g_j  \int\limits_{(m_i+m_j)^2} ds\sqrt{s}
K_1(\sqrt{s}/T) p_{ij}^2\sigma_{ij}(s)}
                         {2T\big(\sum\limits_i g_i m_i^2 K_2(m_i/T)\big)^2 }\;,
\label{sigmav}
\end{equation}
where $\sigma_{ij}$ is the total cross section for annihilation of
a pair of supersymmetric particles into some Standard Model
particles, and  $p_{ij}$ is the momentum of the incoming
particles in their center-of-mass frame.
\beqn
p_{ij}=
\frac{1}{2}\left[\frac{(s-(m_i+m_j)^2)(s-(m_i-m_j)^2)}{s}\right]^\frac{1}{2}
\eeqn
 The summation is over all
supersymmetric particles. Integrating Eq.~\ref{dydt} from $T=\infty$ to
$T=T_0$ would lead $Y_0$.

\subsection{Freeze-out and  approximate solution}
Although one can solve for $Y$ numerically, the procedure is extremely
time consuming especially when scanning over a large parameter
space and when we include a great number of processes. It is
therefore important to seek as  good an approximation as possible
to speed up the code. We will follow the usual procedure of defining 
a freeze-out temperature $T_f$\cite{GondoloGelmini}. 
At high $T$, the LSP are very close to
equilibrium and thus $Y\simeq Y_{eq}$. This will  hold until
freeze-out, where $Y$ will be almost constant whereas $Y_{eq}$
will decrease significantly. 
At high $T$,  one can make the approximation that
$d(Y-Y_{eq})/dT$ is negligible. The freeze-out temperature $T_f$
can be defined from $Y_f=Y(T_f)=(1+\delta)Y_{eq}(T_f)$ with
$\delta $ some (small) constant number, $T_f$  can then be
extracted by solving
\begin{equation}
\label{freeze-out}
\frac{dln(Y_{eq})}{dT}=\sqrt{\frac{\pi g_*(T)}{45G}}<\sigma v>
Y_{eq}\delta(\delta+2) \label{fzot}
\end{equation}
 In the second regime, where $Y \gg Y_{eq}$, one can neglect $Y_{eq}^2$
completely. One finds\cite{GondoloGelmini}
\beqn
\frac{1}{Y(0)}=\frac{1}{Y_f}+\sqrt{\frac{\pi}{45G}}\int_{T_0}^{T_f}
g_*^{1/2}(T) <\sigma v> dT\;\;, \label{Y0eq}
\eeqn

In our code, we in fact integrate from $T=0$ rather than $T=T_0$.
This above approach to finding an approximation was used in
numerous papers \cite{EU,GLP}. However one usually omits the
$1/Y_f$ term but choose $\delta=.5$. The authors of
ref.\cite{GondoloGelmini} state that $Y_0$ as given in
(\ref{Y0eq}) with  $\delta=1.5$  agrees better with the precise
numerical solution. 
We checked this statement for a wide set of
MSSM parameters and reached the same conclusion. 
The approximation agrees with the
 exact result of a full numerical solution within 2\%
\footnote{ The widely used approximation without the $1/Y_f$ term
in Eq.~\ref{Y0eq} and $\delta=0.5$ sometimes gives an error of about
10\%.}.
 Furthermore we find  that the solution
(\ref{Y0eq}) does not depend significantly on $\delta$, varying 
$\delta$ in the range $1<\delta<2$ only affects the result by at most $1\%$.

It is convenient to express the freeze-out temperature $T_f$ in term
of 
\beqn
X_f=\frac{m_{\chi_1^0}}{T_f}
\label{Xfdef}
\eeqn
Typically $X_f\approx 25$, this provides a good initial value for the iterative
solution of Eq.~\ref{fzot}.  
Only a few iterations are necessary to find a solution that satisfies   
 Eq.~\ref{fzot} with $\delta=1.5\pm.2$. 

Typical  freeze-out temperatures vary  between  1GeV  and 10GeV,
in this temperature range, the
$h_{eff}$ and $g_*$ functions can  vary by about 20\%
\cite{GondoloGelmini}. Thus, if one needs an accuracy better than
10\%, one  cannot use a constant value for these functions as it
is done in some papers \cite{GLP}. In our program we  use the
numerical tables of the {\tt  DarkSUSY} package \cite{Darksusy} for a precise
evaluation of $h_{eff}$ and $g_*$.

\subsection{Numerical integration and summation}

  In order to find $T_f$ by solving Eq.~\ref{fzot},  we have to evaluate 
several integrals and  perform a summation over different annihilation channels.
In the evaluation of the thermally averaged cross-section we have included all
two-body subprocesses involving two LSP's, the LSP and a 
co-annihilating 
SUSY particle (all the  particles of the MSSM) as well as 
all subprocesses involving two coannihilating SUSY particles. 
The final states include all possible standard model and Higgs particles that contribute to
a given  process at tree-level.
 The total
number of processes exceeds 2800 not including charged conjugate processes.
In practice  processes involving the heavier SUSY particles contribute only 
when there is a near mass degeneracy with the LSP
since there is a strong Boltzmann
suppression factor $B_f$ in Eq.~\ref{sigmav}. 
\begin{equation}
\label{beps}
B_f=\frac{K_1((m_i+m_j)/T_f)}{K_1(2m_{\tilde{\chi}_1^0}/T_f)}\approx
e^{-X_f\frac{(m_{i}+m_{j}-2m_{\tilde{\chi}_1^0})}{m_{\tilde{\chi}_1^0}}}
\end{equation}
where $m_{i},m_{j}$ are the masses of the incoming particles.
 To speed up the program a given  subprocess  is removed from the sum
(\ref{sigmav}) if the total mass
of the incoming particles is such that $B_f$  is below some limit, 
 $B_\epsilon$  defined by the user.
In most cases  the cross-section for a 
subprocess with coannihilation  is of the same order as the one 
for  the main subprocess with only the LSP,
$\sigma_{coan}\approx\sigma_{\tilde{\chi}_1^0\tilde{\chi}_1^0}$.
To give a precision of $1\%$ it would be sufficient to use $B_\epsilon=10^{-2}$.
However, 
  in our calculation we use a more restrictive value,
$B_\epsilon=10^{-6}$, to be safe and to allow for cases where  
$\sigma_{coan}\gg \sigma_{\tilde{\chi}_1^0\tilde{\chi}_1^0}$.
This can occur for example for coannihilation processes with squarks which depend on
$\alpha_s$,
for processes with poles or in some regions of parameter space where 
$\sigma_{\tilde{\chi}_1^0\tilde{\chi}_1^0}$ is suppressed.
This  criteria, in the case $X_f=25$,  corresponds to a restriction on 
the masses  of the  incoming  particles 
\begin{equation}
m_i+m_j < 2.55\; m_{\tilde{\chi}_1^0}
\end{equation}

  In our program we  provide two options to do the integrations,
the {\it fast} one and the {\it accurate} one. The  {\it fast} mode already gives
a   precision of about 1\% which is good enough for all practical purposes. 
The  {\it accurate}  mode should be used only for some checks.
In the  {\it accurate}  mode   the program evaluates all integrals by means of
an adaptative Simpson  program. It automatically detects all singularities of
the integrands and  checks the precision.  
In the case of the {\it fast} mode the accuracy  is not checked. We integrate the 
squared matrix elements over $\cos\theta$, the scattering angle, by means of
a 5 point Gauss formula. For integration over $s$, Eq.~\ref{sigmav}
 we use a restricted set
of points which depends whether we are in the vicinity of 
a s-channel  Higgs(Z,W) resonance or not. We increase the number of poins if
the Boltzmann factor corresponding to  $m_{pole}$ 
is larger  than $ 0.01\cdot B_\epsilon$.

\subsection{Higgs width}
When the LSP is near a Higgs resonance, they annihilate very efficiently. 
The value of the neutralino annihilation cross-section  depends on the total
width
which can be very large at Higgs masses of 1TeV or so. This is particularly so
at large $\tan\beta$ due to the enhancement in  the $\bbar$ channel. However  
the width of $h(H,A)\ra b\bar{b}$ receives important QCD corrections. Typically 
for the heavy Higgses($>1$TeV) the partial width into $q\bar{q}$ can vary easily by a
factor of 2 from the tree-level prediction, due mostly to the running of the quark mass at high scale.
To take these corrections into account we have redefined the vertices
 $hq\overline{q},Hq\overline{q}$ and $Aq\overline{q}$   using an effective mass that reproduces the
 radiatively corrected width. 
 We used  {\tt HDECAY}\cite{HDECAY} to produce  a table of mass-dependent
 QCD-corrected Higgs partial widths.\footnote{We have neglected the SUSY decay modes of the Higgses. These
 are  small in the cases where the contribution of the Higgs channel to the relic density 
 is very important,
 that is when $m_H\approx 2m_{LSP}$.}
 From this, the effective quark masses $m_b(m_H)$
 are extracted and simple interpolation 
 can reproduce the one-loop corrected width for any value of 
  $m_H$. 
In the region of physical interest, the precision on the width is at the
 per-mil level safe for
  the neutral Higgses partial widths near the $\ttbar$ 
  threshold. However, this region does not contribute significantly to the
 neutralino  cross-section. For the charged Higgs, one can extract from {\tt 
 HDECAY}
 both an effective $m_t$  as well as an effective $m_b$ using high and 
 low $tan\beta$ values.
  This way we could reproduce at 
  better than 1\% level the partial width $H^+\ra t\bar{b}$. 
  The effective quark mass $m_b(Q)$ evaluated at $Q=m_1+m_2$, the sum of the initial masses,
   is used by default
in the $(h,H,A)-\bbar$ vertices. This will lead to the correct result 
when the contribution of the Higgs resonance is very important ($M_H\approx
2m_{\tilde{\chi}_1^0}$).

\subsection{Convention for MSSM parameters}

The input parameters are the ones of the soft SUSY Lagrangian defined at the
weak scale, using the same notation as in {\tt CompHEP/SUSY} models\cite{SusyComphep}. 
In the model used, the masses of fermions of the first generation
are set to zero. The masses of the quarks of the second generation
are also set to zero, so that there is no  mixing of the squarks of the first 
two generations. However we have kept the mass for the muon as well as  the 
trilinear coupling $A_\mu$ as this is relevant for the calculation of the
muon anomalous magnetic moment. 
Although first generation sleptons are
pure left/right states, they are properly ordered according to their masses so
that the correct coannihilating particle corresponding to the lightest slepton
is taken into account.  In order to avoid unnecessary confusion in 
the sign convention for the parameters $\mu$ and $A$, 
we give explicitly the mass matrices for SUSY particles.

The physical masses for charginos and neutralinos are obtained after
diagonalizing the mass matrices, for charginos:

\beqn
{\cal M_C}=\left(\begin{array}{cc}
M_2 & \sqrt{2}M_W\sin\beta\\
\sqrt{2}M_W\cos\beta &   \mu     \\
\end{array}\right)
\eeqn

\noi
and for neutralinos
\beqn
{\cal M_N}=\left(\begin{array}{cccc}
M_1 & 0 & -M_Z\cos\beta s_W & M_Z\sin\beta s_W\\
0 & M_2 & M_Z\cos\beta c_W & -M_Z\sin\beta c_W\\
-M_Z\cos\beta s_W& M_Z\cos\beta c_W&  0 & -\mu \\
M_Z\sin\beta s_W & -M_Z\sin\beta c_W& -\mu & 0 \\
\end{array}
\right)
\eeqn

For sfermions,  the  mixing becomes significant  only for the
third generation, for t-squark, the mass matrix reads

\beqn
{\cal M}^2_{\tilde{t}}=\left(\begin{array}{cc}
m_{\tilde{Q}_{3L}}^2+m_t^2+M_Z^2\cos 2\beta(\frac{1}{2}-\frac{2}{3}\sww) & m_t(A_t-\mu\cot\beta)\\
m_t(A_t-\mu\cot\beta) &   m_{\tilde t_{R}}^2+m_t^2+\frac{2}{3}M_Z^2\cos 2\beta\sww
   \\
\end{array}
\right)
\label{sfermionmass}
\eeqn

while for b-squarks 
\beqn
{\cal M}^2_{\tilde{b}}=\left(\begin{array}{cc}
m_{\tilde{Q}_{3L}}^2+m_b^2+M_Z^2\cos 2\beta(-\frac{1}{2}+\frac{1}{3}\sww) & m_b(A_b-\mu\tan\beta)\\
m_b(A_b-\mu\tan\beta) &   m_{\tilde {b}_R}^2+m_b^2-\frac{1}{3}M_Z^2\cos 2\beta\sww
   \\
\end{array}
\right)
\eeqn
and staus,
\beqn
{\cal M}^2_{\tilde{\tau}}=\left(\begin{array}{cc}
m_{\tilde{L3}_L}^2+m_\tau^2+M_Z^2\cos 2\beta(-\frac{1}{2}+\sww) & m_\tau(A_\tau-\mu\tan\beta)\\
m_\tau(A_\tau-\mu\tan\beta) &   m_{\tilde {\tau}_R}^2+m_\tau^2-M_Z^2\cos 2\beta\sww
   \\
\end{array}
\right)
\eeqn

\section{Contents of the \micro~Package}

\micro~ is a C program that also calls some external FORTRAN functions.
\micro\\ relies on {\tt CompHEP}\cite{comphep} for the definition of the parameters and the
evaluation of all cross-sections.
In the package three  sample main programs are provided. The first
uses an input file to read the weak scale soft SUSY parameters, the last two
give the user  the option of working in the context of a SUGRA-type model.
In the latter case, a sample file containing a
 call to {\tt ISASUGRA}\footnote{The link to Isajet is done here for Version 7.58.} to run the renormalization group equations from the GUT
scale to  the weak scale is provided.
This assumes that the user has already installed  {\tt ISASUGRA} independently
\cite{ ISASUGRA}.
The user has the choice to substitute his/her own favourite RGE code.
\micro~ is a self-contained program with all subroutines included
safe for the {\tt ISASUGRA} routines.
The package includes routines for the input of the MSSM/SUGRA parameters, calls to {\tt CompHEP} to
calculate physical parameters, calculation of $\Omega h^2$ as well
as  calls to other subroutines that check for other constraints on the
model such as $b\ra s\gamma$,$(g-2)_\mu$ as well
as direct limits on the masses of the sparticles. 

\begin{table*}[htb]
\caption{\label{inputfile}Weak scale input parameters in the notation of {\tt CompHEP}}
\begin{tabular}{ll}
\hline
mu  &$\mu$ Higgs mass parameter \\ 
MG1 &$M_1$ Gaugino mass\\
MG2 & $M_2$ Gaugino mass\\ 
MG3 & $M_3$ Gaugino mass\\    
Ml1 & Left-handed slepton mass, 1st generation, $m_{\tilde{L}_{1L}}$\\
Ml2 & Left-handed slepton mass, 2nd generation, $m_{\tilde{L}_{2L}}$\\
Ml3 & Left-handed slepton mass, 3rd generation, $m_{\tilde{L}_{3L}}$\\
Mr1 & Right-handed selectron mass, $m_{\tilde{e}_R}$ \\ 
Mr2 & Right-handed smuon mass, $m_{\tilde{\mu}_R}$\\ 
Mr3 & Right-handed stau mass, $m_{\tilde{\tau}_R}$\\ 
Mq1 & Left-handed squark mass, 1st generation, $m_{\tilde{Q}_{1L}}$\\
Mq2 & Left-handed squark mass, 1st generation, $m_{\tilde{Q}_{2L}}$\\ 
Mq3 & Left-handed squark mass, 1st generation, $m_{\tilde{Q}_{3L}}$\\     
Mu1 & Right-handed u-squark mass, $m_{\tilde{u}_R}$\\
Mu2 & Right-handed c-squark mass, $m_{\tilde{c}_R}$\\
Mu3 & Right-handed t-squark mass, $m_{\tilde{t}_R}$\\  
Md1 & Right-handed d-squark mass, 1st generation$m_{\tilde{d}_R}$\\
Md2 & Right-handed s-squark mass, $m_{\tilde{s}_R}$\\ 
Md3 & Right-handed b-squark mass, $m_{\tilde{b}_R}$\\ 
Atop & Trilinear coupling for top quark, $A_t$\\   
Ab & Trilinear coupling for b-quark, $A_b$\\   
Atau& Trilinear coupling for $\tau$-lepton, $A_\tau$\\
Am& Trilinear coupling for $\mu$, $A_\mu$\\   
MH3 & Mass of Pseudoscalar Higgs ($M_A$)\\   
tb  & Ratio of the vacuum expectation values of the scalar doublets \\   
\hline
\end{tabular}
\end{table*}

Some test input files  with the MSSM weak scale
parameters are stored in the {\tt data/} and {\tt datap/} directories together with the reference
output file for comparison purposes.

\subsection{Input parameters}

\noi{\bf Weak scale parameters}

When running \micro, the input file
  must specify the  set of weak scale 
 SUSY parameters as in Table~\ref{inputfile}.
 The sfermion masses correspond to the soft masses without the  D-term. 
The  convention for the third generation of sfermions is specified explicitly 
 in Eq.~\ref{sfermionmass}.
As in  the {\tt CompHEP/SUSY} model\cite{SusyComphep}, only the mass matrices for the
third generation will then be diagonalized to obtain the physical masses.

\begin{table*}[htb]
\caption{\label{sm}
Standard Model Parameters}
\vspace{.5cm}
\begin{tabular}{lll}
\hline\hline
EE    &0.31223      &Electromagnetic coupling constant\\
GG    &1.238        &Strong coupling constant at $M_z$ \\
SW    &0.473        &sin of the Weinberg angle, $\sin\theta_W$ \\
s12   &0.221        &sin of Cabibbo angle \\
MZ    &91.1884      &Z mass\\
Mm    &0.1057       &mass of muon\\
Mtau    &1.777        &mass of tau-lepton\\
Mc    &1.42         &c-quark mass\\
Mtop  &175          &top quark mass\\
Mb    &4.62         &b-quark mass\\
wtop  &1.7524       &width of top quark\\
wZ    &2.4944       &Z-boson width\\
wW    &2.08895      &W-boson width\\
\hline\hline
\end{tabular}
\end{table*}

In addition to these parameters one can choose to redefine parameters, such as
the  standard model parameters (Table~\ref{sm}),
that are fixed by default. One can also redefine SUSY parameters that do not
appear in Table 1, such as the widths of Higgses or SUSY particles.
For example, in processes with t-channel 
poles it is sometimes necessary to specify a width for
some particles such  as gauginos. By default these widths have been set to 1GeV.
Even though the LSP is assumed to be stable, to avoid any spurious pole it is 
necessary to introduce also a small width. Numerically, the results do not
depend on the exact value chosen for these widths.  
A complete list of widths for
all supersymmetric particles can be found in Table~\ref{list}.

\begin{table}
\caption{\label{list}Names, masses and widths of
supersymmetric particles in {\tt CompHEP/SUSY} notation}
\begin{tabular}{|l|l|l|l||l|l|l|l|}
\hline
 Name     &symbol&mass&width   &  Name     &symbol&mass&width  \\
\hline
chargino 1   &\~{}1+&MC1   &wC1   & mu-sneutrino &\~{}nm&MSnmu &wSnmu \\ 
chargino 2   &\~{}2+&MC2   &wC2   & tau-sneutrino&\~{}nl&MSntau&wSntau\\
neutralino 1 &\~{}o1&MNE1  &wNE1  & u-squark L   &\~{}uL&MSuL  &wSu1  \\
neutralino 2 &\~{}o2&MNE2  &wNE2  & u-squark R   &\~{}uR&MSuR  &wSu2  \\
neutralino 3 &\~{}o3&MNE3  &wNE3  & c-squark 1   &\~{}c1&MSc1  &wSc1  \\
neutralino 4 &\~{}o4&MNE4  &wNE4  & c-squark 2   &\~{}c2&MSc2  &wSc2  \\
gluino       &\~{}g &MSG   &wSG   & t-squark 1   &\~{}t1&MStop1&wStop1\\
selectron 1  &\~{}e1&MSe1  &wSe1  & t-squark 2   &\~{}t2&MStop2&wStop2\\
selectron 2  &\~{}e2&MSe2  &wSe2  & d-squark L   &\~{}dR&MSdL  &wSd1  \\
smuon 1      &\~{}m1&MSmu1 &wSmu1 & d-squark R   &\~{}dL&MSdR  &wSd2  \\
smuon 2      &\~{}m2&MSmu2 &wSmu2 & s-squark L   &\~{}sL&MSsL  &wSs1  \\
stau 1       &\~{}l1&MStau1&wStau1& s-squark R   &\~{}sR&MSsR  &wSs2  \\
stau 2       &\~{}l2&MStau2&wStau2& b-squark 1   &\~{}b1&MSbot1&wSbot1\\
e-sneutrino  &\~{}ne&MSne  &wSne  & b-squark 2   &\~{}b2&MSbot2&wSbot2\\
\hline
\end{tabular}
\end{table}

\begin{table}
\caption{\label{Higgs}Parameters for Higgs particles in {\tt CompHEP/SUSY} notation}
\begin{tabular}{|l|l|l|l||l|l|l|l|}
\hline
 Name     &symbol&mass&width   &  Name     &symbol&mass&width  \\
\hline
Light Higgs  &h  &Mh    &wh    & CP-odd Higgs &H3 &MH3   &wH3   \\
Heavy higgs  &H  &MHH   &wHh   & Charged Higgs&H+ &MHc   &wHc   \\
\hline
\end{tabular}
\end{table}

\noi{\bf SUGRA GUT-scale Parameters}

The value of the weak scale soft supersymmetric Lagrangian can be extracted
from {\tt ISASUGRA}. 
To use this option  it is  necessary to specify the path for the \isajet libraries.
The only difference between this option and the weak scale parameters
we have just mentionned
is that the {\tt readInitFile} has been  replaced 
and an additional call to the {\tt sugraVar} function is done (see the description
of these functions in Section 3.2). 
 
 The set of input  parameters for the  SUGRA model are the usual
\begin{verbatim}
m0      common scalar mass at GUT scale
mhf     common gaugino mass at GUT scale
a0      trilinear soft breaking parameter at GUT scale
tb      tan(beta) or the  ratio of vacuum expectation values 
sgn     +/-1,  sign of Higgsino mass term
mtop    pole mass of the top quark
model    = 1  SUGRA model
         = 2  SUGRA with true gauge coupling unification at the GUT scale
\end{verbatim}

\subsection{Functions of \micro}

All functions provided with the package are briefly described below.
The  user can find examples of how to call these routines in the
 {\tt omg.c}  or {\tt sugomg.c} file  located in the 
root directory.  The types of functions called are specified in Table
\ref{functiontype}.

\begin{table*}[htb]
\caption{\label{functiontype}Main functions called}
\vspace{.5cm}
\begin{tabular}{l}
\hline
{\tt double darkOmega(double *Xf,int fast, double Beps)}\\
{\tt int readInitFile(char * filename)}\\
{\tt int assignVal(char * name, double val)}\\
{\tt int  sugraVar(float m0,float mhf,float a0,float tb,float sgn, 
float mtop,}\\{\tt int model)}\\
{\tt int calcDep(int LC, int Widths)}\\
{\tt int findVal(char * name, double * val)}\\
{\tt void printMasses(FILE * f, int sort)}\\
{\tt char * lsp(void)}\\
{\tt void printMainChannels(FILE *f, double cut)}\\
{\tt void printMSSMvar(FILE *f)}\\
{\tt double delrho(void)}\\
{\tt double gmuon(void)}\\
{\tt int MassLimits(void)}\\
{\tt double bsganew(void)}\\
\hline
\end{tabular}
\end{table*}

\noi
$\bullet$ {\tt darkOmega(Xf,fast,Beps)}

\noi
This is the basic function of the package 
which returns the relic density $\Omega h^2$ (Eq.~\ref{omegah}). {\tt Xf} is the initial 
value of  the $X_f$ parameter needed for solving iteratively the 
freeze-out temperature, Eq.~(\ref{freeze-out}). We recommend  to use {\tt Xf}$= 25$ for the 
first call, this is the value set by default in the sample main file.
  The   value of {\tt Xf} returned by {\tt  darkOmega}   corresponds to the
 solution of  the freeze-out equation (\ref{freeze-out}). 
If the user  calls {\tt darkOmega} in  a scan where 
the model parameters are modified slowly,  we recommend  to use the preceding {\tt Xf} value 
as input for the next iteration. Otherwise, it is better to assign to {\tt Xf} some fixed 
value before each call. 
The parameter {\tt Beps} defines the criteria for including a given channel in the
sum for the calculation of the thermally averaged cross-section,
Eq.~(\ref{beps}), 
$10^{-6}$ is the  recommended value. 

If  {\tt fast $\neq$ 0}, the program uses the   {\it fast} mode, otherwise 
the  {\it accurate} mode is used, see Section 2.2.  
If some problem is encountered, {\tt darkOmega} returns -1.

\noi
$\bullet$ {\tt readInitFile(filename)}

\noi
Reads the input file, {\tt filename}. The input parameters of the MSSM 
that need to be specified 
are given in Table \ref{inputfile}. 
 The function returns: \\
\hspace*{1cm}  {\tt 0} - then the input has been read correctly;\\
\hspace*{1cm}  {\tt -1} - if the file does not exist 
or can not be opened for reading;\\ 
\hspace*{1cm}  {\tt n} - when  the line number {\tt n} has been written in the
wrong format.\\
The correct format of a line is \\
\hspace*{2cm}{\tt <name of variable><space><numerical value>} \\
for example \\ 
\hspace*{2cm} {\tt MG3 1500.} \\
This function writes a WARNING if one of the variables of Table 1 is
not defined in {\tt filename}

\noi
$\bullet${\tt assignVal(name,val)}

\noi
Changes the value of one of the standard model or MSSM parameters.
{\tt name} is either one of the parameter names  given in Tables~\ref{inputfile},\ref{sm}
or a width of a particle whose name is specified in Tables~\ref{list},\ref{Higgs}. 
{\tt val} is the new value. 
The function returns 0 when it successfully recognizes
the parameter name  and 1 otherwise. The only
reason for an  error is  a wrong  parameter name.

\noi
$\bullet${\tt calcDep(LC,Widths)}

\noi
 Calculates all parameters of the MSSM model including physical masses and mixings as
 well as couplings. The calculation of the Higgs masses is based on
 {\tt FeynHiggsfast}\cite{FeynHiggs}.
 
{\tt calcDep} returns a non zero error code when it is unable to calculate some physical
parameter. For example, this can occur when the input parameters are such 
 that one ends up with a negative squared mass for a sfermion.
 The  input parameters of {\tt calcDep} should be set to {\tt LC=1} and {\tt Width=1}. 
We provide this option only to give the possibility to make comparisons with other programs.
 {\tt LC=0}  
switches off the QCD loop correction for the Higgs bosons  decays to {\tt b,c},and {\tt t}
 quarks. 
 {\tt Width=0} switches off 
the automatic calculation of widths of Higgs bosons. If {\tt Width=0}  the user 
has to define the   Higgs  width values  by means of either  {\tt readInitFile} or 
{\tt assignVal}. The names of the corresponding variables  are 
listed in Table \ref{Higgs}.  
Note that the choice for the {\tt Width} parameter affects only the
Higgs width in the propagator. The value of the Hqq vertex is calculated
independently either at the tree-level(LC=0) or one-loop level(LC=1) and
could be different from
the one that reproduces the Higgs total width. 
    
We have to   emphasize that  the user must call {\tt calcDep}  after having 
defined the 
 free MSSM parameters and before the calculation of the relic density. Otherwise 
the calculated relic density is not valid.

\noi
$\bullet${\tt findVal(name,\&val)}

\noi
Outputs the value of a parameter of the MSSM, either one of the input parameters
or one of the derived parameters. 
{\tt name} should be the name of a variable (free  
or constrained one).  {\tt findVal} returns zero if the name indeed corresponds 
to some variable and {\tt 1} otherwise. 
For a name that has been correctly specified, {\tt findVal}  assigns to {\tt val} the 
numerical  value of the parameter. This function is particularly useful if one wants to impose
constraints on the masses of the superparticles. The
names of the  masses of superparticles and Higgses together with the names of
sparticles in {\tt CompHEP} notation can be found in Tables~\ref{list},\ref{Higgs}
\footnote{Note that some of the mass parameters can be negative,   
the  masses of particles are automatically redefined such that they 
correspond  to the absolute values  for these parameters.}.

\noi
$\bullet${\tt printMasses(filename, sort)}

This routine prints the masses of the supersymmetric particles mentionned
in Table~\ref{list} as well as all Higgs masses and widths into the specified file {\tt filename}. 
To see the masses on the
 screen replace {\tt filename} by {\tt stdout}. 
 If sort$\neq 0$ the masses are sorted in order of increasing mass.

\noi
$\bullet${\tt printMSSMvar(filename)}   
  
Prints the  MSSM parameters presented in the Table 1.
To see the parameters on the
 screen replace {\tt filename} by {\tt stdout}. 

\noi
$\bullet${\tt printMainChannels(filename,cut)}

Allows the user to see which channels 
dominate in the calculation of $\Omega h^2$.   
This function  writes into the file {\tt filename},
the channels whose relative contribution to $1/\Omega h^2$ exceeds the value chosen for {\tt cut}. 
The value of the
contribution in percent is also written. If {\tt cut=0}, all channels will be displayed. 

\noi
$\bullet${\tt sugraVar(m0,mhf,a0,tb,sgn,mtop,model)}

\noi
This function  solves the renormalization group equations
and assigns a value to all the MSSM weak scale parameters in Table 1,
except for the
trilinear coupling for the muon, {\tt Am} ($A_\mu$).
We have used the approximation,
$A_{\mu}=A_0-.7*M_{1/2}$ to define this parameter.	
 This function  calls the {\tt Isajet} package, which must be installed 
 by the user.
 {\tt  sugraVar} returns zero for a succesful assignment of weak scale
 parameters or  the \isajet error code for a wrong set of parameters.
 The error code corresponds to the NOGOOD  parameter of  COMMON /SUGPAS/, 
 (see the \isajet manual\cite{ISASUGRA}).
 To substitute another RGE code, the user must modify the
 files sources/sugrac.c.
  
\noi
$\bullet${\tt lsp()}   
  
This function returns the name of the LSP.
The relic density can be calculated with any particle being the LSP.
If the user wants to impose a specific LSP,  the nature of the LSP
must be checked after calling  {\tt calcDep}. In the sample {\tt omg.c} file
that is provided in the package,
this is done for the conventional choice of a neutralino LSP.

\noi
$\bullet${\tt bsganew()}

Returns the value of the branching ratio for  $b\ra s\gamma$.
For $b\ra s\gamma$ we have improved on the results of \cite{bsgamma}
by including some very
recent new contributions beyond the leading order that are
especially important for high $\tan\beta$. We include these terms by
following \cite{bsgammaNLO}. This function should be called after all parameters of the
MSSM have been defined in {\tt calcDep}.

\noi
$\bullet${\tt gmuon()}

Returns the value of the supersymmetric contribution to 
the anomalous magnetic moment of the muon\cite{g-2,g-2nous}. 
The result depends only on  the parameters of the gaugino mass matrices
as well as on the smuon parameters, in particular $A_\mu$ the
trilinear coupling. This parameter has been added in the function 
{\tt sugraVar} described above as this is not calculated by
\isasugra.

\noi
$\bullet${\tt delrho()}

Computes the stop/sbottom contribution
to the $\Delta\rho$ parameter, including the two-loop
gluon exchange and pure scalar diagrams\cite{deltarho}.
Returns the value for $\Delta\rho$ and a WARNING if the value exceeds
the  limit $\Delta\rho>1.3\times 10^{-3}$.
This routine is included in {\tt FeynHiggsFast}\cite{FeynHiggs}.

\noi
$\bullet${\tt MassLimits()}

This function returns a value greater than 1 and 
 prints a WARNING when the choice of parameters conflicts with a
direct accelerator limits on sparticle masses.
Only the direct limits  from
LEP in the general MSSM are implemented. In constrained models more severe
limits can be set, this should be added by the user.
The constraint on the light Higgs mass is not implemented and must be
added by the user.

\section{Compilation and installation instructions.}

The package can be obtained at\\ 
{\tt http://wwwlapp.in2p3.fr/lapth/micromegas}

\noi
After unpacking the files, the root directory of the package,
\begin{verbatim} micromegas_1.0 \end{verbatim}   
will be created. 
This directory contains a makefile, three sample main programs 
{\tt omg.c, sugomg.c} and {\tt scycle.c} and
two directories for data files.
The compiler options and the location of the \isajet libraries, when needed,
should be defined in the makefile. The compiler options for
Linux, Alpha and Silicon Graphics
workstations  are defined automatically.
At the moment the package cannot be compiled on HP Workstations.

To compile,  type\\
\hspace*{2cm}{\tt make }\\
This creates the 
{\tt omg} executable file for the omg.c main program
which  does not require the \isajet library.
If {\tt omg}  is launched  without an argument  it calculates the relic density 
for the default parameters which corresponds to the data/datap1 file  MSSM.
The output is presented in Section 5.
 
If {\tt omg} is launched with some arguments, it treats all of them as 
names of  input files and calculates  the relic densities 
for each of them. Say,\\ 
\hspace*{2cm} {\tt            omg datap/* }\\
produces the results presented in the Table \ref{comparison}.

As the compilation of the code for  more than 2800 processes would take a long time
while in most cases only a small fraction of processes are needed, we have
generated the processes and linked them   dynamically
 only when they are needed. 
The first time  new processes need to be linked
compilation can take up to a few minutes, the actual computation time remains 
insignificant.

The {\tt sugomg.c} code contains an example of application of 
the {\tt sugraVar} routine. It can be compiled by \\
\hspace*{2cm} {\tt           make main=sugomg }\\
The corresponding executable file is {\tt sugomg}. It is necessary 
to first install the \isajet package. The  environment variable SUGRA is used
to define the path to the {\tt isajet.a}  library. This variable should be 
defined properly by modifying  the {\tt makefile}.

A sample file {\tt scycle.c} for making a scan over two parameters ($m_0,m_{1/2}$)
in the SUGRA model is also provided. The range of values for  $m_0,m_{1/2}$
need to  be specified, the remaining parameters have to be given as inputs.
It can be compiled by \\
\hspace*{2cm} {\tt           make main=scycle }\\
The corresponding executable file is {\tt scycle}.

To install the package one needs initially about 10MB of disk space.
As the program generates libraries for annihilation processes only at the
time they are required, the total disk space necessary can 
increase significantly after running the program for different models.
These libraries are stored in the {\tt sources/2-2/} subdirectory
and can be removed by the user. However, this is done  
at the expense of slowing down the compilation whenever the 
program will need again  these libraries.

\section{Sample Output}

Running \micro~ with the default value of the standard parameters and with
the input file datap/datap1, which contains the parameters of the MSSM as specified in
 Table~\ref{comparison}  
model A, will produce the following output:

\begin{verbatim}
Initial file  "data/datap1"

Higgs masses and widths
h   : Mh    =  125.1 (wh    =3.2E-03)
H   : MHH   = 1000.8 (wHh   =2.3E+00)
H3  : MH3   = 1000.0 (wH3   =2.4E+00)
H+  : MHc   = 1003.1 (wHc   =2.3E+00)

Masses of SuperParticles:
~o1 : MNE1  =    22.4 || ~ne : MSne  =    80.0 || ~nm : MSnmu =    80.0 
~l1 : MStau1=    92.7 || ~nl : MSntau=    93.1 || ~m1 : MSmu1 =   108.0 
~e1 : MSe1  =   108.2 || ~e2 : MSe2  =   111.2 || ~m2 : MSmu2 =   111.3 
~l2 : MStau2=   124.7 || ~1+ : MC1   =   198.6 || ~o2 : MNE2  =   200.1 
~o3 : MNE3  =   270.3 || ~g  : MSG   =   300.0 || ~2+ : MC2   =   332.4 
~o4 : MNE4  =   332.6 || ~t1 : MStop1=   813.0 || ~b1 : MSbot1=   998.4 
~c1 : MSc1  =   998.7 || ~uL : MSuL  =   998.7 || ~uR : MSuR  =   999.4 
~c2 : MSc2  =   999.4 || ~dR : MSdR  =  1000.3 || ~sR : MSsR  =  1000.3 
~dL : MSdL  =  1001.6 || ~sL : MSsL  =  1001.6 || ~b2 : MSbot2=  1003.5 
~t2 : MStop2=  1175.8 || 

Channels which contribute more than 1 %.
 25% ~o1 ~o1 -> e E 
 25% ~o1 ~o1 -> m M 
 39% ~o1 ~o1 -> l L 
  4% ~o1 ~o1 -> ne Ne 
  4% ~o1 ~o1 -> nm Nm 
  2% ~o1 ~o1 -> nl Nl 
Omega=2.48E-01
g2=2.55E-09
bsgamma=4.4E-04
\end{verbatim}

Running \micro~ in the SUGRA case, with {\tt ISASUGRA} version 7.58
and the input parameters 
\begin{verbatim}
sugomg 100 400 0 10 +1 175 1
\end{verbatim}
\noi
will write the values of the input parameters, the value
of the weak scale parameters as calculated by {\tt ISASUGRA}, the corresponding
list of weak scale parameters in {\tt CompHEP} notation followed 
by the usual output of \micro~ as  described above:

\begin{verbatim}
 MSSM parameters:
mu      513.940063
MG1     162.922958
MG2     311.489258
MG3     914.497253
Ml1     286.277618
Ml2     286.277618
Ml3     285.412537
Mr1     178.365540
Mr2     178.365540
Mr3     175.411743
Atau    -244.407349
Am      -280.000000
Mq1     819.445862
Mq2     819.445862
Mq3     760.147278
Mu1     793.780884
Mu2     793.780884
Mu3     659.573975
Md1     790.153137
Md2     790.153137
Md3     785.846619
Atop    -723.817383
Ab      -1038.150269
MH3     576.585266

Higgs masses and widths
h   : Mh    =  115.2 (wh    =3.2E-03)
H   : MHH   =  576.9 (wHh   =1.2E+00)
H3  : MH3   =  576.6 (wH3   =1.3E+00)
H+  : MHc   =  582.2 (wHc   =1.3E+00)

Masses of SuperParticles:
~o1 : MNE1  =   160.8 || ~l1 : MStau1=   175.7 || ~m1 : MSmu1 =   183.4 
~e1 : MSe1  =   183.4 || ~nl : MSntau=   278.2 || ~ne : MSne  =   279.1 
~nm : MSnmu =   279.1 || ~e2 : MSe2  =   290.2 || ~m2 : MSmu2 =   290.2 
~l2 : MStau2=   292.3 || ~1+ : MC1   =   297.0 || ~o2 : MNE2  =   297.3 
~o3 : MNE3  =   518.2 || ~o4 : MNE4  =   534.5 || ~2+ : MC2   =   534.6 
~t1 : MStop1=   623.2 || ~b1 : MSbot1=   752.2 || ~dR : MSdR  =   790.5 
~sR : MSsR  =   790.5 || ~uR : MSuR  =   793.0 || ~c1 : MSc1  =   793.0 
~b2 : MSbot2=   796.1 || ~uL : MSuL  =   817.7 || ~c2 : MSc2  =   817.7 
~t2 : MStop2=   818.3 || ~dL : MSdL  =   821.6 || ~sL : MSsL  =   821.6 
~g  : MSG   =   914.5 || 

Channels which contribute more than 1 %.
  1% ~o1 ~o1 -> b B 
 17% ~o1 ~o1 -> e E 
 17% ~o1 ~o1 -> m M 
 18% ~o1 ~o1 -> l L 
  5% ~o1 ~l1 -> Z l 
 20% ~o1 ~l1 -> A l 
  1% ~o1 ~m1 -> Z m 
  5% ~o1 ~m1 -> A m 
  1% ~o1 ~e1 -> Z e 
  4% ~o1 ~e1 -> A e 
  3% ~l1 ~l1 -> l l 
Omega=2.07E-01
g2=1.46E-09
bsgamma=3.48E-04
\end{verbatim}

\section{Results and Comparisons}
 
The \micro~ code was extensively  tested against another public package 
for   calculating  the  relic density, {\tt DarkSUSY}.  
As discussed previously, the two codes  differ somewhat in the numerical method used for 
solving the density equations.
\micro~ includes 
more subprocesses(e.g. all coannihilations with sfermions), 
loop-corrected Higgs widths, and complete tree-level matrix
elements for all processes. {\tt DarkSUSY} includes on the other hand
 some loop induced processes such as $\chi\chi\ra
g g,\gamma\gamma$ which are generally small. 
Whenever the coannihilation channels with sfermions and the Higgs pole are
not important we expect good agreement with {\tt DarkSUSY}.
We have first compared \micro~ with a version of {\tt DarkSUSY} where we
have replaced the matrix elements by {\tt CompHEP} matrix elements. 
As expected,  we found 
excellent numerical agreement between the two programs
(at the $2\%$ level) for all
points tested. We have then made 
comparisons with the original
version of {\tt DarkSUSY}. 
The results of these comparisons 
  are displayed for a few test points in Table~\ref{comparison}. 
We found in general good agreement between the two codes. 
However we have observed some discrepancies that could reach up to  ($30\%$) 
in particular in
the process $\chi_1^0\chi_1^+\ra t\overline{b}$(model B in Table~\ref{comparison})
\footnote{The \comphep~ result for this matrix element agrees with the result of 
{\tt GraceSUSY}\cite{gracesusy}.}.
 As displayed in the line $\Omega_{tree}^{\chi}$,
when removing non-gaugino coannihilation channels and reverting to the
tree-level treatment of the Higgs width we recover results similar to 
{\tt DarkSUSY}.
The impact of these extra channels, model C for sleptons and
model D for squarks can be as large as an order of magnitude and depends
critically on the mass difference with the lightest neutralino.

\begin{table*}[htb]
\caption{\label{comparison}Sample results and comparison with \darksusy}
\begin{tabular}{|l|l|l|l|l|l|l|}
\hline
Model&A  &B &C &D&E&F  \\
\hline
tb  &5.& 10.& 10.&10. &45.& 50.\\   
mu  &264.5  &400.&518.6 & -1200.&500.& 1800.\\
MG1 &25.9 & 500.& 166.1&300.&180.&850.  \\
MG2 &258.9 &1000. & 317.9&600.& 350.& 1600.\\
MG3 &800. &3000. &931.8 & 1800.&1000.&4000. \\
\hline
Ml1 &101. & 1000.& 289.0&1000.& 500.& 2000.\\
Ml3 &112. & 1000.& 288.1&1000.&500.& 2000. \\
Mr1 &100. & 1000.& 177.0&1000.& 300.& 1600.\\ 
Mr3 & 88.  &1000.&174.1 & 1000.&250.& 1600.\\ 
Mq1 &1000. & 1000.& 834.1& 1000.&1000.& 3500.\\
Mq3 &1000.&1000.& 773.9 &1000.& 1000.& 3500.\\   
Mu1 &1000. & 1000.& 803.1& 500.&1000.& 3500.\\
Mu3 &1000. &1000.& 671.8 & 500.&1000.& 3500.\\
Md1 &1000. &1000.& 799.3 & 500.&1000.& 3500.\\
Md3 &1000. &1000.& 799.7 & 500.&1000.& 3500.\\ 
\hline
Atop &2400.  & 0.&-738.1&-1800.& -1000.& -3000.\\\   
Ab &2400.  &0. &-1058.&-1800. &-1000.& -3000.\\  
Atau &0. & 0.& -249.2&0.& -100.& -500.\\ 
Mh3 &1000.  &1000.& 581.9 &1000.&500.& 1700.  \\ 
\hline 
\hline
$\mneuto$ &22.4 & 384.3& 164. & 299.9 &178.2 & 849.3 \\
$\mneutt$ &200.3 &402.7 &303.5 &597.9 &333.9 &1585.4 \\
$\mchargo$ &198.6 & 395.5& 303.3& 597.9& 333.8& 1585.3\\
\hline
$m_{\tilde{e}_1}$ &108.2 & 1000.9& 182.1&1000.9 & 303.0& 1600.6 \\
$m_{\tilde{\tau}_1}$ &92.7 & 997.5& 174.4& 990.3& 236.9& 1595.0\\ 
$m_{\tilde{t}_1}$ &813.0 & 1007.7& 635.6& 326.8& 930.7& 3439.9\\ 
\hline
$m_h$ &125.1  & 111.2& 115.3& 118.4& 118.1 & 121.7\\
\hline
$g-2(\times 10^{-10})$&25.5 &1.88 &13.9 &-1.66 &30.4 &2.07\\
$b\ra s\gamma(\times 10^{-4})$ &4.4 &3.8 &3.5 &4.9 &2.8 &3.5\\
\hline
\hline
{\tt micrOMEGAs}&.25  &.024 &.14&.15&.22&.26 \\ 
$\Omega_{tree}^{\chi}$&.25 &.024&.32 &.74&.13&.56  \\ 
\hline 
\darksusy&.25  &.018 &.32&.74&.13& .55\\
\hline 
\end{tabular}
\end{table*}

The effect of the Higgs width is particularly important at large $\tan\beta$
with the enhanced contribution of the $b$-quark. However 
the one-loop correction  amounts to a reduced
effective b-quark mass and a much smaller width especially at large values of
$m_H$. If it was not for the strong
Boltzmann suppression factor singling out the contribution at $\sqrt{s}\approx
2 m_\chi$ there will be little difference after integrating  over the peak for
the one-loop or tree-level result. However the effect observed can be as much as a
factor 2. For $m_\chi\approx M_A/2$ the narrower
resonance suffers less from the Boltzmann reduction factor leading to
$<\sigma^{1-loop}>/<\sigma_{tree}> > 1$ and $\Omega_{1-loop}<\Omega_{tree}$ (model F).
Further away from the pole however one catches the contribution from the wider
resonance without excessive damping from the Boltzmann factor, as
expected  $\Omega_{1-loop}>\Omega_{tree}$ (model E).

Our numerical results were also compared with Ref. \cite{benchmark}.
Qualitative agreement is found in the case of SUGRA model although we use a
different RGE code. Precise comparisons necessitates a careful tuning of
parameters to make sure we have the same parameters at the weak scale. 
A random scan over $m_0-m_{1/2}$ for $\mu>0$ shows the typical shape of the
allowed region ($.1<\Omega h^2<.3$) in SUGRA model for moderate values of $\tan\beta$.

\begin{figure*}[htb]
\begin{center}
\caption{$\Omega h^2$ in the $m_0-m_{1/2}$ plane in a SUGRA model
with $\tan\beta=10$, $\mu>0$ and $m_{top}=175$GeV. The dark (blue)
region corresponds to $.1< \Omega h^2< .3$ and the light (pink)
region to $\Omega h^2< .1$. In the white region at large $m_0-m_{1/2}$,
the relic density is above the present limit $\Omega>.3$ and
 in the region at small $m_0$, the 
$\tilde\tau$ is the LSP. The yellow (light grey) region correspond to the region
 excluded by $b\ra s\gamma$. The LEP limit on $m_h=113$GeV is also displayed.}
\mbox{\epsfxsize=16cm\epsfysize=12cm\epsffile{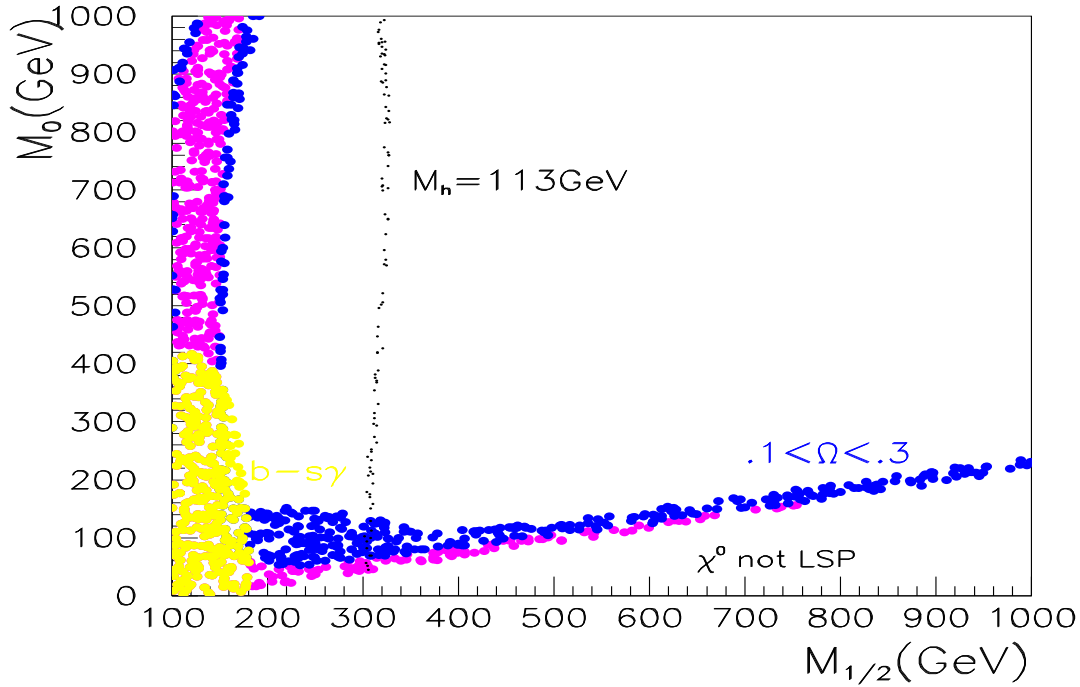}}
\vspace*{-1.5cm}
\end{center}
\end{figure*}

\section{Conclusion}
The package \micro~ allows to calculate the relic density of the LSP in the most general
MSSM with $R_P$ conservation. The package is self-contained safe for the 
{\tt ISASUGRA} 
package that is required when
using the SUGRA option. All possible channels for coannihilations are 
included and all matrix elements are
calculated exactly at tree level with the help of {\tt CompHEP}. 
Loop corrections for the masses of Higgs particles (two-loop) and the width of the
Higgs (QCD one-loop) are implemented. 
Good agreement with existing calculations is found when identical set of channels
are included.
Future versions will  include 
loop corrections to neutralino masses. Even though these corrections are only
a few GeV's they can alter significantly the calculation of the relic density 
when there is a near mass degeneracy with the next to lightest supersymmetric particle
that contributes to a coannihilation channel \cite{benchmark}.
Although  the loop processes are in general  small, we will include   
 $\chi\chi\ra\gamma\gamma,\gamma Z, gg$  in an update of \micro.

\section{Acknowledgements}

This work was supported in part by the PICS-397, {\it Calcul en physique des particules}. 
The work of A.~Semenov and A.~Pukhov was supported in part by the 
 CERN-INTAS grant
99-0377 and by RFFR grant 01-02-16710.
The work of A. Pukhov was also supported by the 
grant, Universities of Russia 015.02.02.05. We thank A.~Cottrant for providing the
code for the $(g-2)_\mu$.

\end{document}